\documentclass[aps,pra,twocolumn,superscriptaddress,groupedaddress,longbibliography]{revtex4-1}
\usepackage{epsfig,amsopn}
\usepackage{graphicx}
\usepackage{color}
\usepackage{amsmath,amssymb}
\usepackage{enumerate}
\newcommand\bea{\begin{eqnarray}}
\newcommand\eea{\end{eqnarray}}
\newcommand\beq{\begin{equation}}
\newcommand\eeq{\end{equation}}

\def\nn{\nonumber}
\def\f{\frac}

\def\ep{\epsilon}

\def\si{\sigma}

\def\De{\Delta}
\def\dg{\dagger}
\def\la{\langle}
\def\ra{\rangle}

\begin{document}
\title{Majorana fermions in Kitaev chains side-coupled to normal metals} 
\author{ Abhiram Soori }
\email{abhirams@uohyd.ac.in}
\affiliation{ School of Physics, University of Hyderabad, Prof. C. R. Rao Road, Gachibowli, Hyderabad-500046, India.}

\begin{abstract} 
Majorana fermions, exotic particles with potential applications in quantum computing, have garnered significant interest in condensed matter physics. The Kitaev model serves as a fundamental framework for investigating the emergence of Majorana fermions in one-dimensional systems.  We explore the intriguing question of whether Majorana fermions can arise in a normal metal (NM) side-coupled to a Kitaev chain (KC) in the topologically trivial phase. Our findings reveal affirmative evidence, further demonstrating that the KC, when in the topological phase, can induce additional Majorana fermions in the neighboring NM region. Through extensive parameter analysis, we uncover the potential for zero, one, or two pairs of Majorana fermions in a KC side-coupled to an NM. Additionally, we investigate the impact of magnetic flux on the system and calculate the winding number -a topological invariant used to characterize topological phases.
\end{abstract}

\maketitle

\section{ Introduction}
 The Kitaev model stands as a cornerstone in the study of Majorana fermions (MFs) within condensed matter physics~\cite{kitaev2001unpaired}.  This involves a p-wave superconducting term in the Hamiltonian of one-dimensional spinless normal metal (NM). When the superconducting pairing term is nonzero, the open system described by the Hamiltonian is in trivial phase without any MFs or in topological phase with a pair of MFs depending on whether $|\mu|>2t$ or $|\mu|<2t$ respectively, where $\mu$ is the chemical potential and $t>0$ is the hopping strength.  Theoretical proposals for realizing MFs revolutionized the search for these exotic particles~\cite{lutchyn2010majorana,oreg2010helical}. Experimental evidences for MFs were reported~\cite{mourik2012,Das2012}.  A minimal version of the model proposed by Kitaev has been experimentally realized recently~\cite{dvir2023}. Extensions of MFs  to periodically driven systems have been studied~\cite{thaku,soorifmf}. Several transport studies on junctions of a metal with topological superconductors that host MFs have been reported~\cite{seti2015,soori19trans,soori20probing}. MFs are proposed to be used in developing qubits that are building blocks of a quantum computer~\cite{plugge2017}. 

\begin{figure}[htb]
\includegraphics[width=4.5cm]{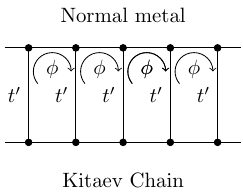}
\caption{Schematic diagram of a Kitaev chain side-coupled to a normal metal. $t'$ is the hopping amplitude between the two chains. $\phi$ quantifies the magnetic flux piercing through each placquette. }\label{fig:schem}
\end{figure}

The topologically trivial phase in Kitaev model is characterized by an insulating phase when the superconducting pairing term is switched off. In other words, when $\mu>2t$ ($\mu<-2t$) the band is completely filled (empty). Hence, turning on superconductivity does not result in MFs. The limit of $\mu<-2t$ corresponds to Bose-Einstein condensate (BEC) superconductors hosting strongly bound pairs when s-wave superconductivity is switched on in spin-half tight binding chain~\cite{chen2005,randeria14,seti2022}. Andreev reflection spectroscopy in such BEC superconductors is expected to show finite subgap conductance when the barrier separating the NM and superconductor is turned into a well~\cite{lewan23}. Physically this can be understood as due to the formation of Andreev bound state due to potential well at the interface.  This means that though the BEC superconductors have no electrons in their NM state, the superconductivity develops in the neighboring metal region if the metal hosts electrons. A similar line of argument can be applied to Kitaev model in the trivial phase to phrase the following question. `Can MFs develop in a NM side-coupled to a Kitaev chain (KC) in the topologically trivial phase?' (See Fig.\ref{fig:schem} for a schematic of the system). This paper answers this question in positive. Further, we find that when KC is in the topological phase, it possible that it can induce another pair of MFs in the neighboring NM region. Depending on the choice of parameters, there can be zero, one or two pairs of MFs in  KC side-coupled to a NM. We consider the possibility of a magnetic flux piercing the region between KC and NM. We calculate winding number - a topological invariant of a KC  side-coupled to NM to characterize the phase of the system. 

\section{ The Hamiltonian and the bulk spectrum }
The Hamiltonian for the system under consideration is: 
\bea 
H  &=& \sum_{n=-\infty}^{\infty}[-t(c^{\dg}_{n+1}c_n+{\rm h.c.})-\mu (c^{\dg}_nc_n-\f{1}{2})\nn \\ &&+\De (c^{\dg}_{n+1}c^{\dg}_n+{\rm h.c.}) -t(f^{\dg}_{n+1}f_n e^{i\phi}+{\rm h.c.}) \nn \\ && - t'(c^{\dg}_n f_n+{\rm h.c.})], \label{eq:ham}
\eea
where $c_n$ ($f_n$) annihilates an electron at site $n$ on KC (NM), $t$ is the hopping strength within KC, $te^{i\phi}$ is the complex hopping amplitude in the NM, $\mu$ is the chemical potential on KC, $\De$ is the pairing strength within KC, $t'$ is the coupling between KC and NM. The phase factor $\phi$ corresponds to a magnetic flux threading through the rectangular plaquette between KC and NM nearest neighbor sites.  
We first convert the Hamiltonian in eq.~\eqref{eq:ham} to momentum space - $H=\sum_k\Psi^{\dag}_k H_k \Psi_k$, where  
\bea 
H_k &=& 
\begin{bmatrix}
-2t\cos{k}-\mu & -t' & -2i\De\sin{k} & 0 \\
-t' & -2t\cos{k'} & 0 & 0 \\
2i\De\sin{k} & 0 & 2t\cos{k}+\mu & t' \\ 
0 & 0 & t' & 2t\cos{k'}
\end{bmatrix}~,\nn \\ && \label{eq:ham-k}
\eea
  $k'=k+\phi$ and $\Psi_k=[c_k, f_k, c^{\dag}_{-k}, f^{\dag}_{-k}]^T$. 
 
\begin{figure}[htb]
\includegraphics[width=8cm]{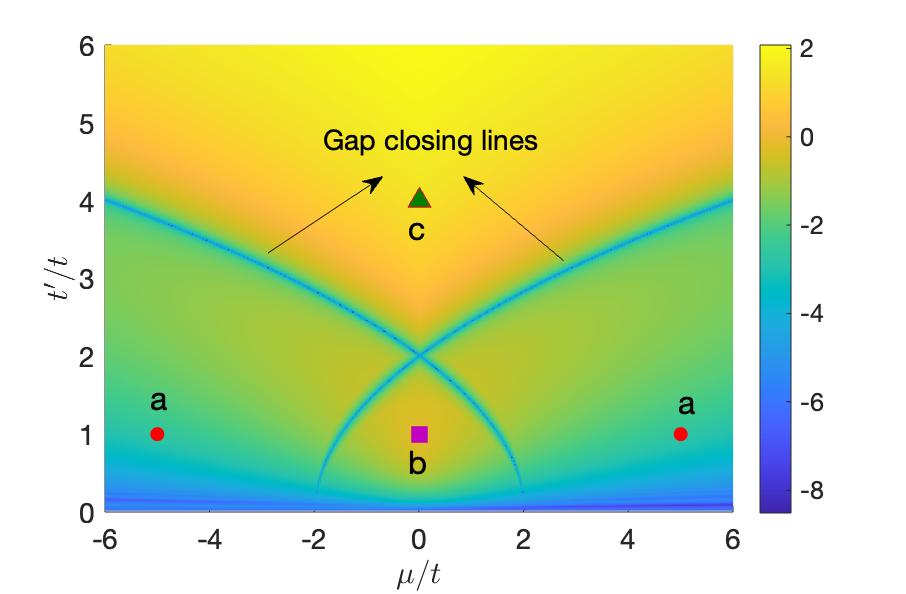}
\caption{Logarithm of the bulk gap in units of $t$ versus $t'$ and $\mu$ for $\De=0.5t$ and $\phi=0$.  The data points with $t'=0$ are excluded. Data points (a) $(\mu,t')=(\pm 5,1)t$, (b) $(\mu,t')=(0,1)t$ and (c) $(\mu,t')=(0,4)t$ are representative points in each of the three gapped regions, for which the spectrum of a finite chain is examined in Fig.~\ref{fig:zero-energy-states}.  }\label{fig:spectrum}
\end{figure}

\begin{figure}[htb]
\includegraphics[width=2.7cm]{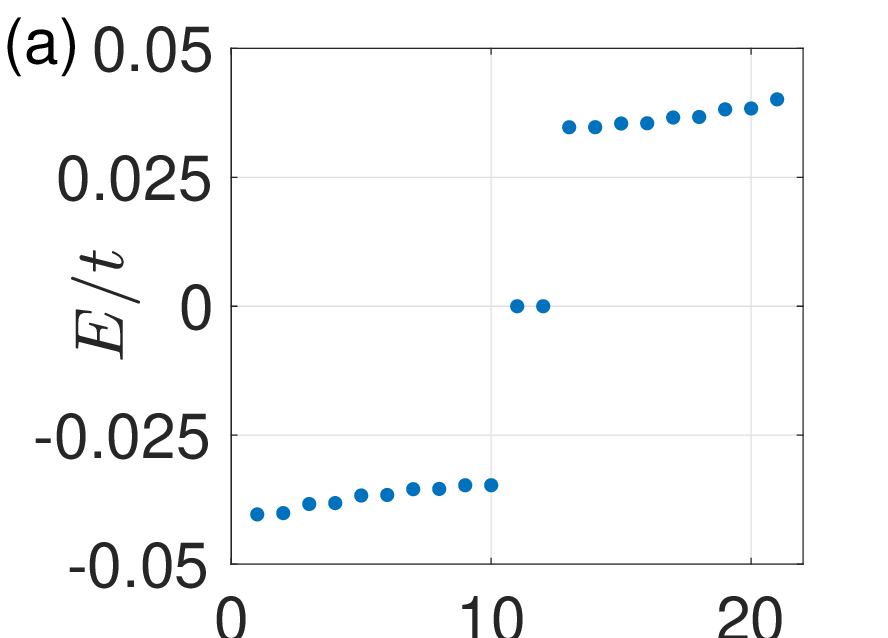}
\includegraphics[width=2.7cm]{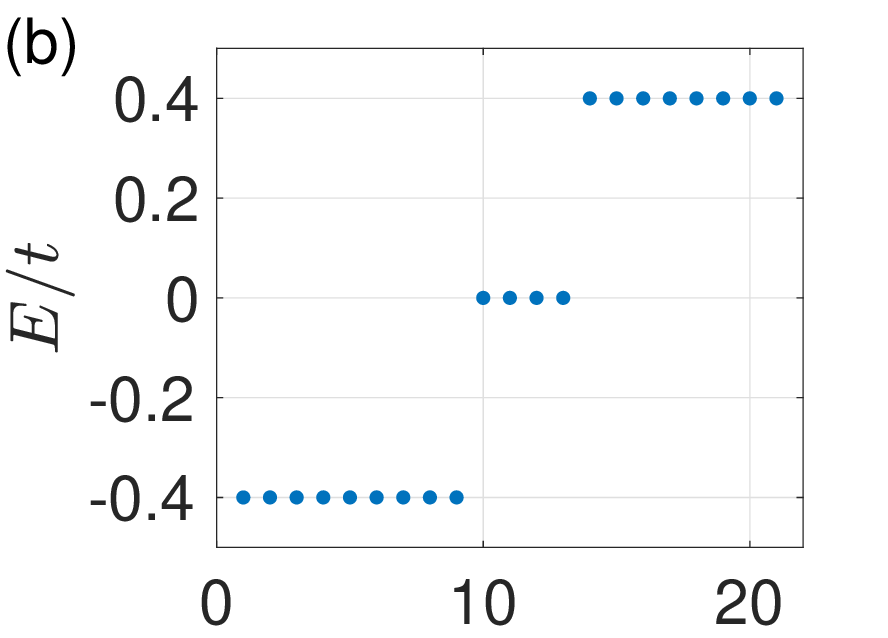}
\includegraphics[width=2.7cm]{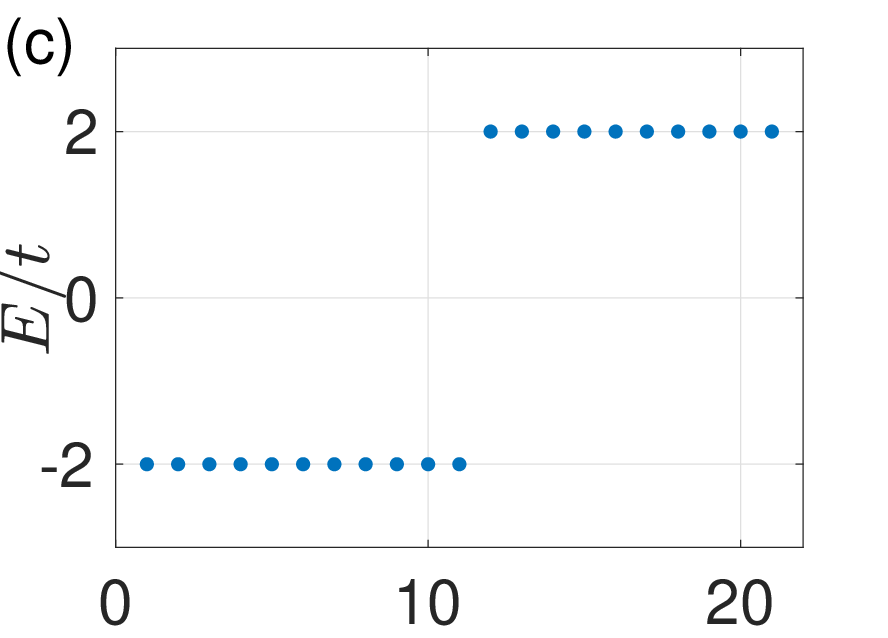}
\caption{Spectrum of finite coupled chains zoomed around zero energy for $L=1500$ sites on each chain on three sides of the gap closing line  with $\De=0.5t$ for: (a) $\mu=\pm 5t$, $t'=t$, (b) $\mu=0$, $t'=t$,  (c) $\mu=0$, $t'=4t$.  Number of pairs of zero energy states: (a) 1, (b) 2, (c) 0. }\label{fig:zero-energy-states}
\end{figure}

Unless mentioned explicitly, $\phi=0$. Taking $\De=0.5t$, we numerically diagonalize the Hamiltonian in eq.~\eqref{eq:ham-k} and plot the logarithm of bulk gap in Fig.~\ref{fig:spectrum}. We find that the bulk gap closes on two curves that intersect at $(\mu,t')=(0,2)t$. In the limit $t'=0$, NM and KC are decoupled and the gap closes trivially since the NM is not gapped.
On three sides of the gap-closing lines, we plot the  spectrum for a finite chain with $L=1500$ sites and zoom it around the zero energy in Fig.~\ref{fig:zero-energy-states} for $\De=0.5t$. We find that for $(\mu,t')=(\pm5t,t)$, there is one pair of zero energy states, for $(\mu,t')=(0,t)$ there are two pairs of zero energy states and for $(\mu,t')=(0,4t)$, there are no zero energy states. Next, we need to check whether the zero energy states found so are MFs and characterize the different sides of the gap closing lines by a topological invariant. 

\begin{figure}[htb]
\includegraphics[width=8cm]{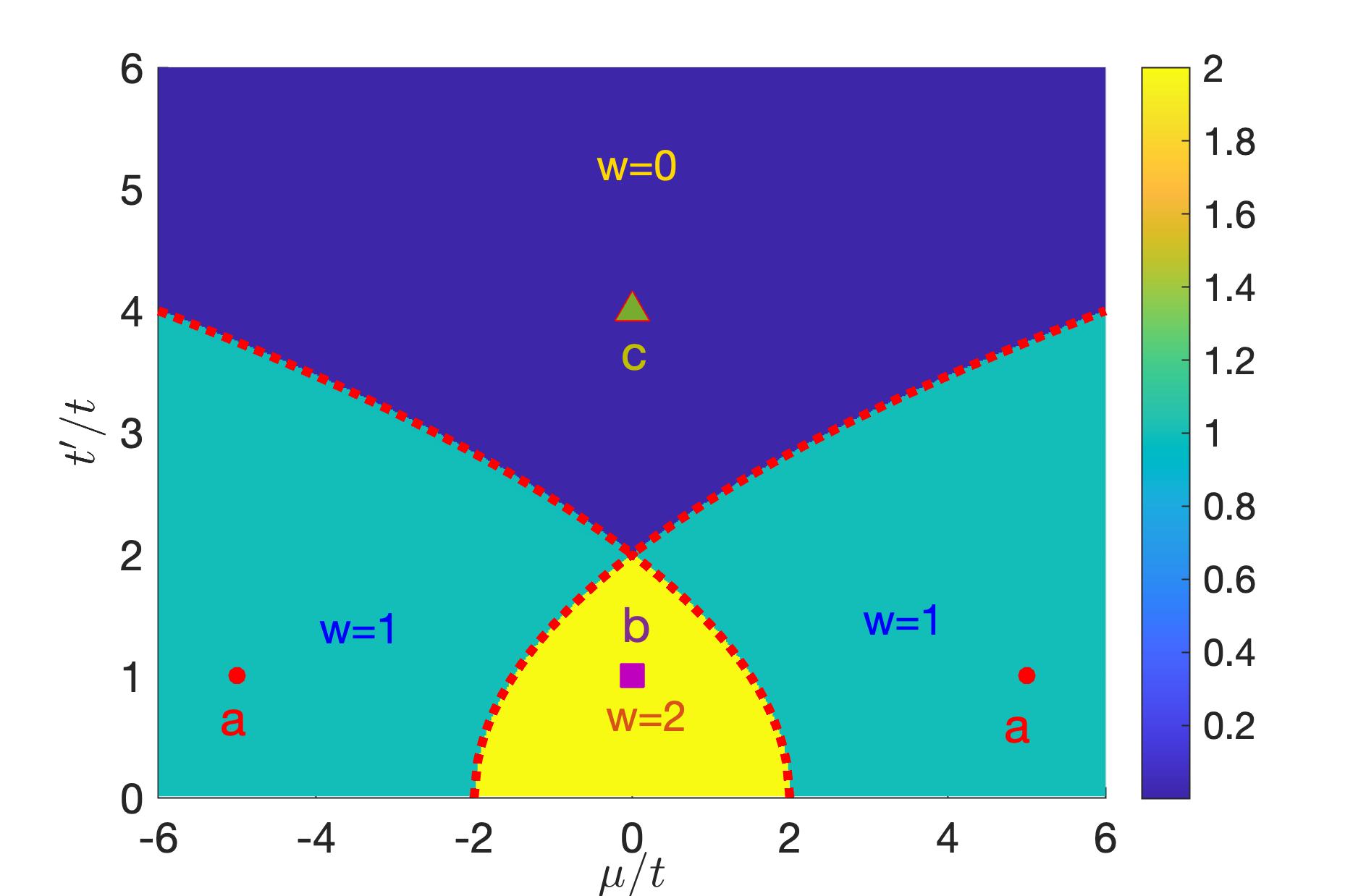}
\caption{Winding number versus $t'$ and $\mu$. Red dotted lines are the lines where $|\mu_{\si}|$ is $2t$. Red dotted lines match with the gap closing lines in Fig.~\ref{fig:spectrum}.  The data points with $t'=0$ are excluded. Data points (a) $(\mu,t')=(\pm 5,1)t$, (b) $(\mu,t')=(0,1)t$ and (c) $(\mu,t')=(0,4)t$ are representative points in each of the three phases, for which the spectrum of a finite chain is examined in Fig.~\ref{fig:zero-energy-states}. Other parameters: $\De=0.5t$, $\phi=0$. }\label{fig:winding}
\end{figure}
\section{ Winding number}
The finding that zero energy states which could possibly be MFs can develop in a hybrid made of two topologically trivial systems is interesting. To investigate whether the zero energy states have Majorana character, we start with the non-superconducting hybrid. The electron eigenenergies of the Hamiltonian eq.~\eqref{eq:ham-k} for $\De=0$, $\phi=0$ are $E=-\mu/2 -2t\cos{k} +\si \sqrt{(\mu/2)^2+t'^2}$, with $\si=\pm 1$. The respective eigenstates are $|\phi^e_{\si}\ra=[\phi_{\si}, 0, 0]^T$ for the electron sector, where $\phi_{\si}$ is the eigenvector of $2\times 2$ block in the electron sector of the Hamiltonian eq.~\eqref{eq:ham-k}. Similarly, the eigenenergies in the hole sector are $E=\mu/2 +2t\cos{k} - \si \sqrt{(\mu/2)^2+t'^2}$, with $|\phi^h_{\si}\ra=[0,0,\phi_{\si}]^T$ being the eigenvectors. The two sets of bands corresponding to $\si=\pm 1$ can be thought of as bonding and anti-bonding orbitals of a two-level system. The effective chemical potential in the sector $\si$ is $\mu_{\si}=\mu/2-\si\sqrt{(\mu/2)^2+t'^2}$. When $\mu_{\si}$ falls outside (inside) the range $(-2t,2t)$, the phase can be expected to be topologically trivial  (nontrivial) in the sector $\si$.

\begin{figure}[htb]
\includegraphics[width=2.7cm]{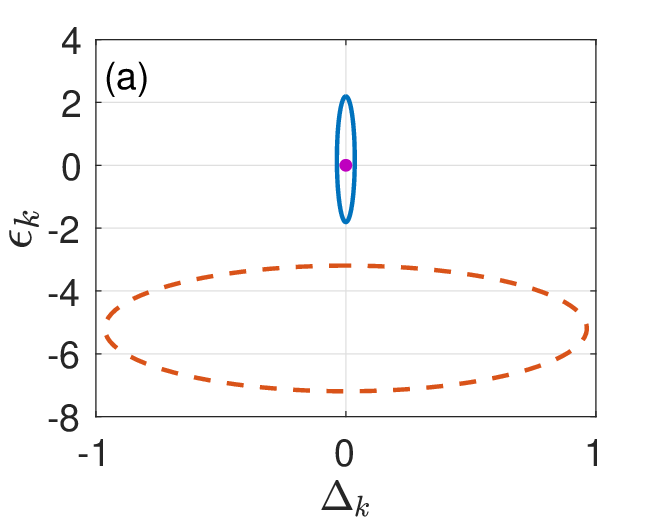}
\includegraphics[width=2.7cm]{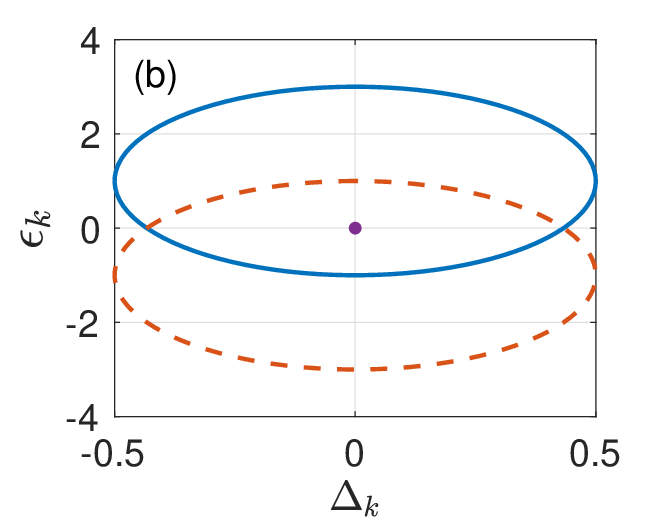}
\includegraphics[width=2.7cm]{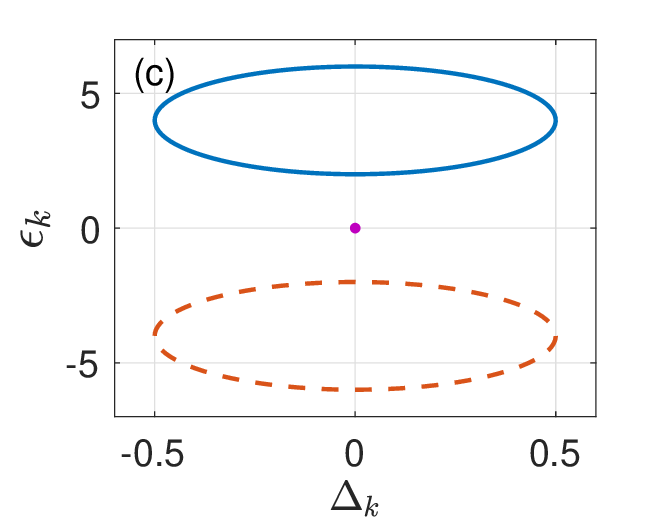}
\caption{ Plots of $\epsilon_k$ versus $\Delta_k$ as $k$ is varied in the range $(0,2\pi]$ for: (a) $\mu=5t$, $t'=t$, (b) $\mu=0$, $t'=t$,  (c) $\mu=0$, $t'=4t$. For all plots, $\De=0.5t$. In each of the subplots, dashed curve corresponds to $\si=-1$ band while solid curve corresponds to $\si=+1$ band. Origin is shown by a magenta dot. In (a), one curve encircles the origin. In (b), both the curves encircle the origin while in (c), both the curves do not encircle the origin. Winding number for (a, b, c) are respectively $(1, 2, 0)$, which agrees with the number of pairs of zero energy modes seen in Fig.~\ref{fig:zero-energy-states}. }\label{fig:windingshow}
\end{figure}

 We project the $4\times 4$ Hamiltonian in eq.~\eqref{eq:ham-k} into the two sectors described by $\si=\pm1$. In sector $\si$, the projected Hamiltonian is given by 
\bea
h^{\si}_k &=& 
\begin{bmatrix} 
\la\phi^e_{\si}|H_k|\phi^e_{\si}\ra & \la\phi^e_{\si}|H_k|\phi^h_{\si}\ra \\ 
\la\phi^h_{\si}|H_k|\phi^e_{\si}\ra & \la\phi^h_{\si}|H_k|\phi^h_{\si} \ra
\end{bmatrix}~.
\eea
Such a matrix has the form $h^{\si}_k = \ep_k\tau_z+\De_k\tau_y$, where $\ep_k$ and $\De_k$ are functions of $k$, and $\tau_z$ and $\tau_y$ are $2\times 2 $ Pauli spin matrices. The winding number in sector $\si$ denoted by $w^{\si}$ is then given by the number of times $(\ep_k,\De_k)$ loops around $(0,0)$ when $k$ is taken from $0$ to $2\pi$. The total winding number $w=w^++w^-$ is a non negative integer which indicates the number of pairs of MFs in the open system. We justify this method used to calculate the winding number in the appendix. For each set of values of the parameters $(t',\mu)$, we numerically find the winding number $w$ and plot it in Fig.~\ref{fig:winding} for $\De=0.5t$. The red dotted lines indicate the values of the parameters for which $|\mu_{\si}|$ crosses $2t$. At the line separating the regions having $w=1$ and $w=2$, one of the two $\mu_{\si}$'s crosses $\pm 2t$. The same is true for the line separating the regions with winding numbers $0$ and $1$.  In Fig.~\ref{fig:windingshow}, we show the plot of $\epsilon_k$ versus $\De_k$ as $k$ is varied in the range $(0,2\pi]$. 

In Fig.~\ref{fig:winding}, we have excluded the data points for $t'=0$, where the gap closes and the winding number is not well defined in one of the sectors. For small values of $t'/t$, the gap is small in the sector $\si=1$. This sector corresponds to the electrons being populated mostly in the NM. Due to the small coupling to KC, superconductivity leaks into NM and a small gap opens up in this sector. This means that the decay length of MFs in the $\si=1$ sector gets larger as $t'/t$ decreases. As $t'/t$ becomes smaller, the length $L$ of the system needed to see two MFs becomes larger. 

\begin{figure}[htb]
\includegraphics[width=4cm]{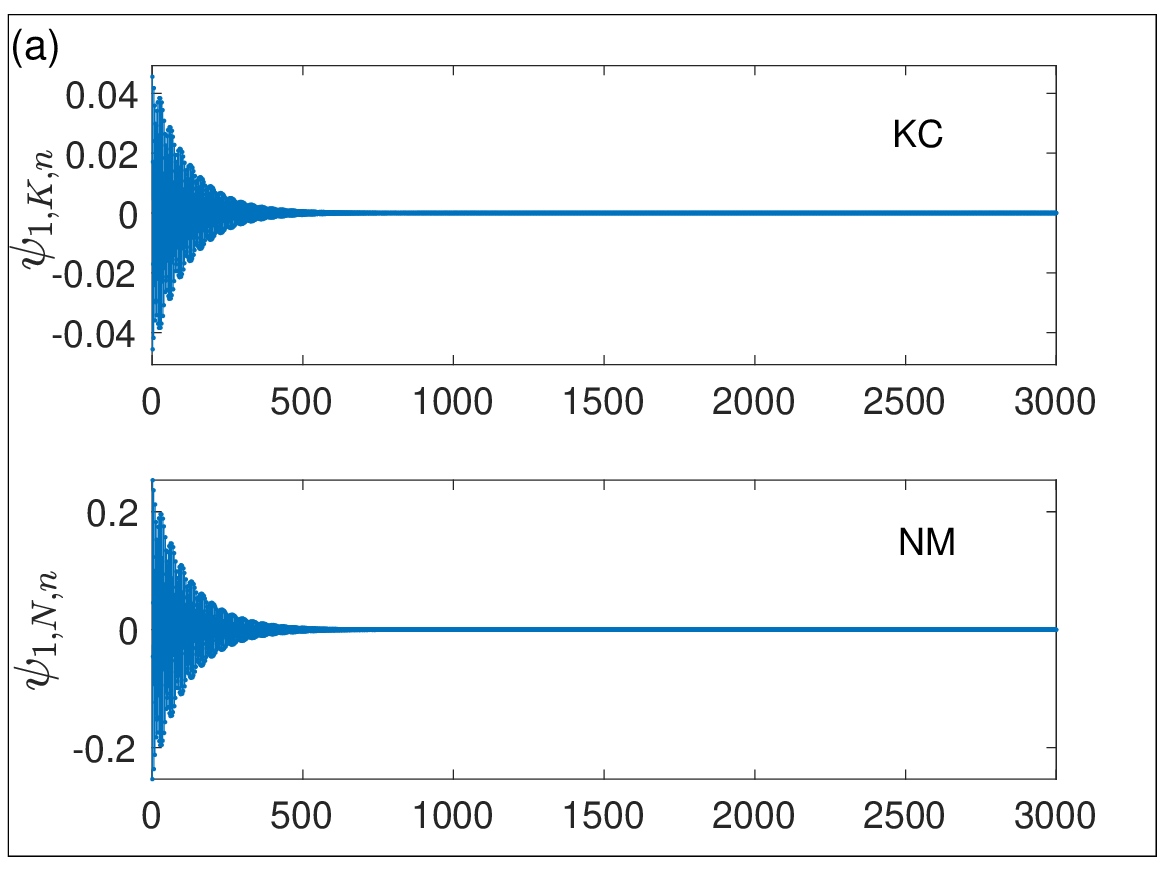}
\includegraphics[width=4cm]{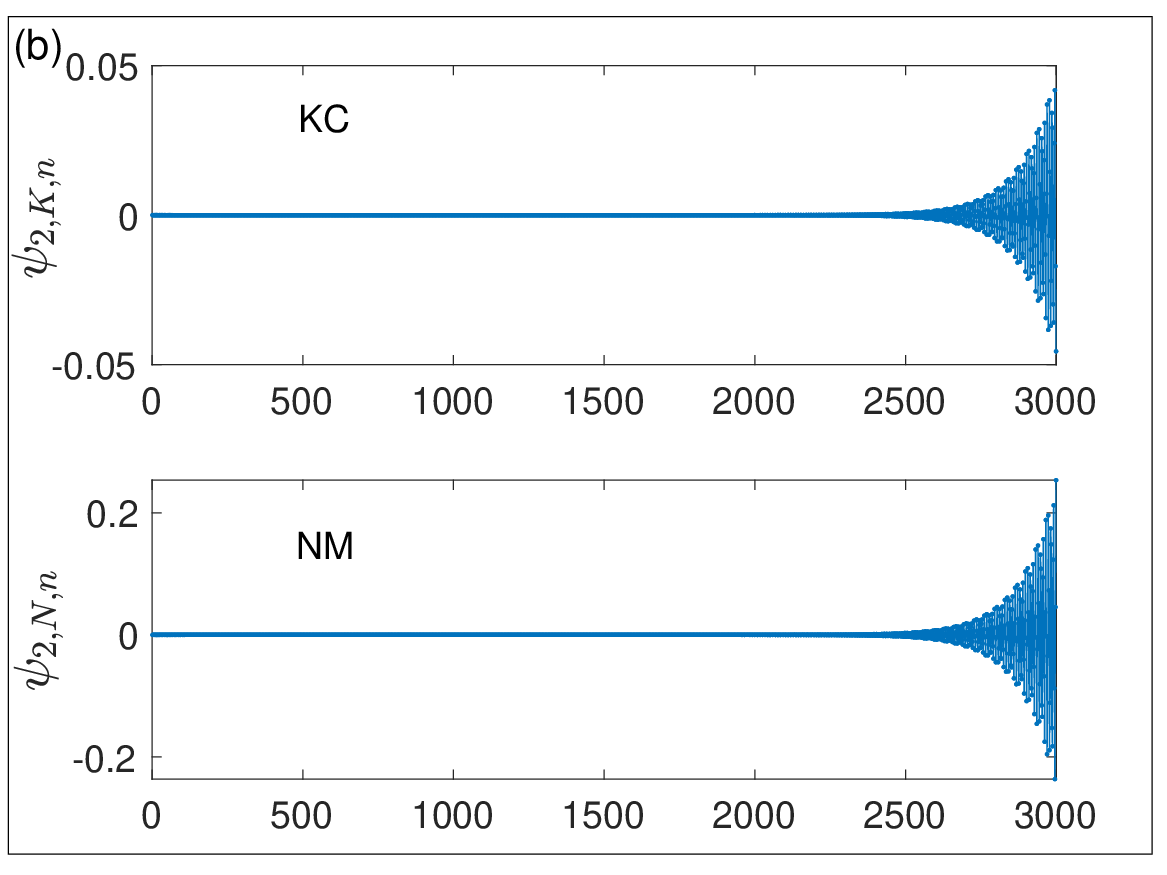}
\caption{Spatial representation of the wavefunctions of the Majorana fermions for $\mu=5t$, $t'=t$, $\De=0.5t$, $L=1500$. Majorana wavefunction is plotted versus site index $n$. $n=2m-1$ ($n=2m$) represents electron (hole) component of the wavefunction located at site $m$. MF wavefunction localized on the (a) left end, (b) right end. The top panel in each sub-figure represents sites on KC and the bottom panel represents NM. The parameters correspond to the winding number $1$. }\label{fig:spatial_w_1}
\end{figure}

\begin{figure}[htb]
\includegraphics[width=4cm]{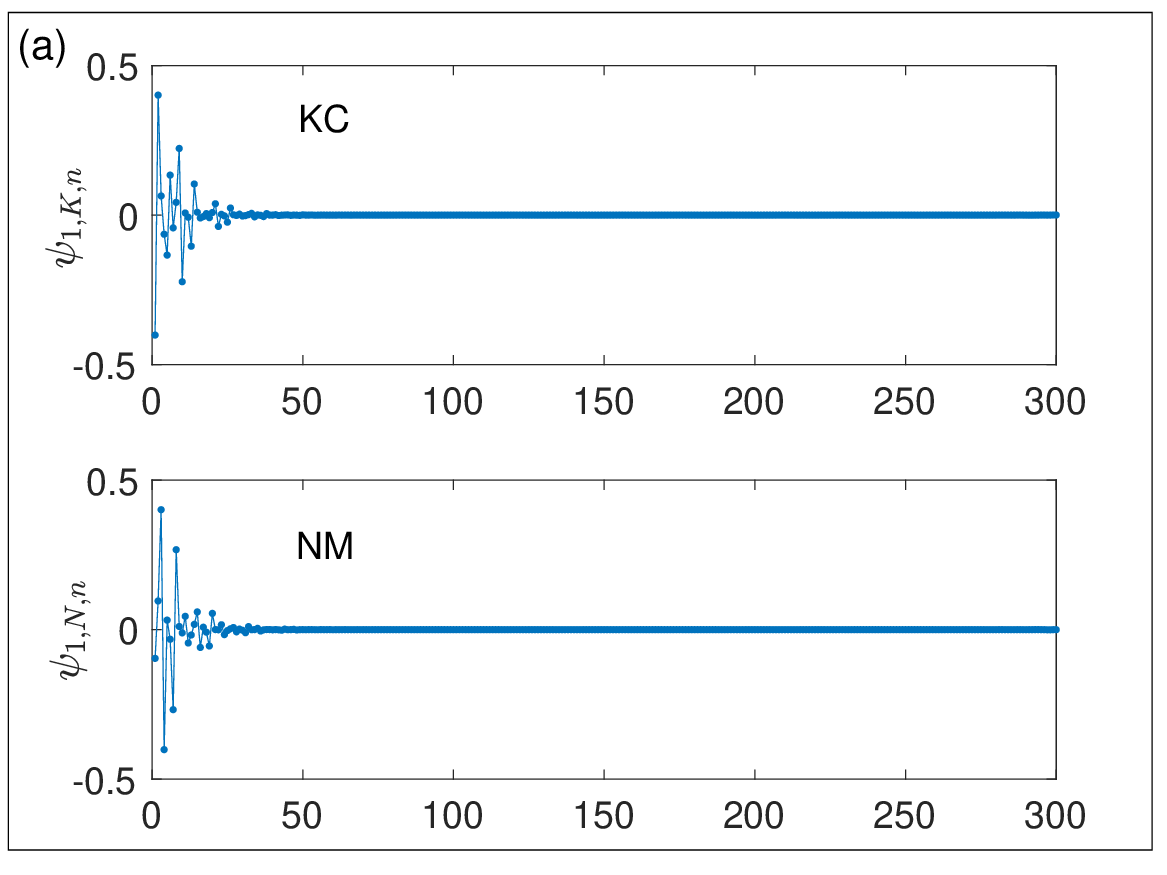}
\includegraphics[width=4cm]{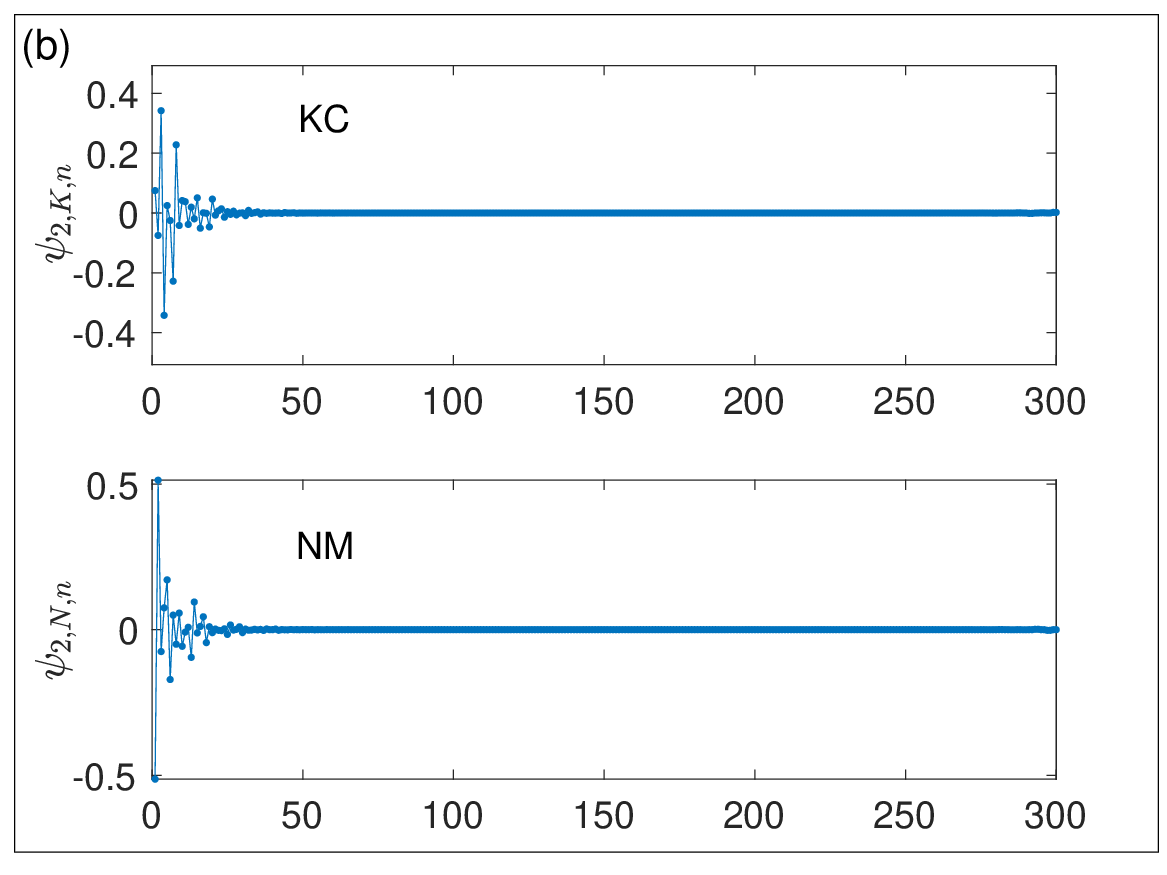}
\includegraphics[width=4cm]{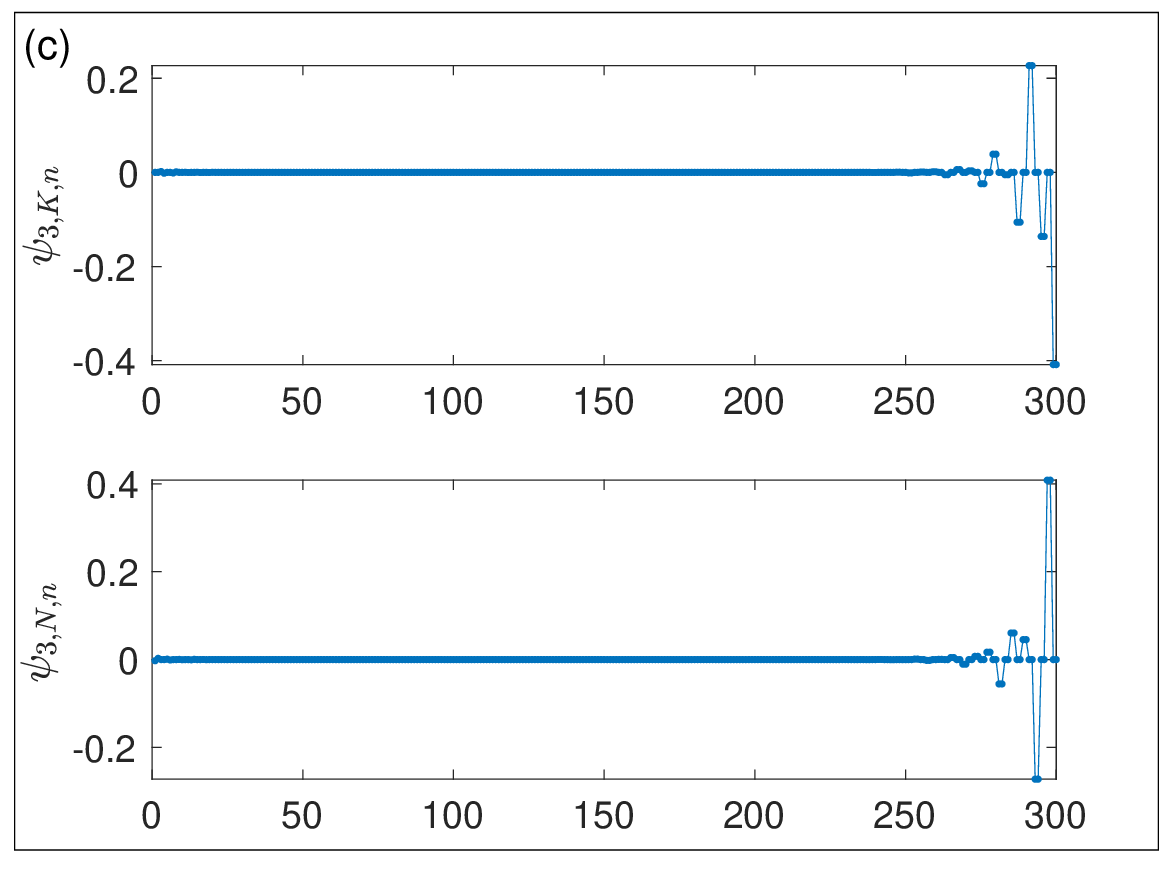}
\includegraphics[width=4cm]{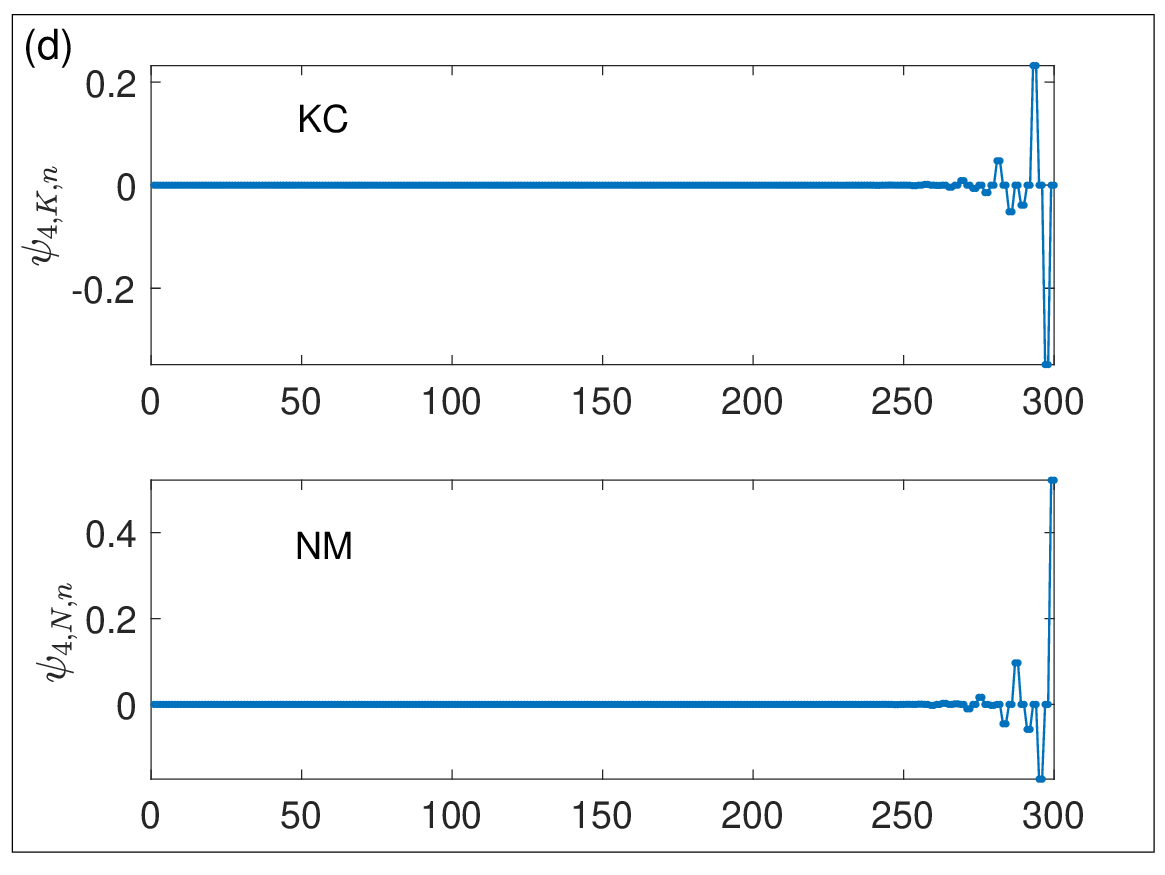}
\caption{Spatial representation of the wavefunctions of the Majorana fermions for $\mu=0$, $t'=t$, $\De=0.5t$, $L=150$. Majorana wavefunction is plotted versus site index $n$. $n=2m-1$ ($n=2m$) represents electron (hole) component of the wavefunction located at site $m$. MF wavefunctions localized on the (a,b) left end, (c,d) right end. The top panel in each sub-figure represents sites on KC and the bottom panel represents NM. The parameters correspond to the winding number $2$. }\label{fig:spatial_w_2}
\end{figure}

We plot the wavefuntions of the MF modes corresponding to data point (a) of Fig.~\ref{fig:winding} [data point (b) of Fig.~\ref{fig:winding}] in Fig.~\ref{fig:spatial_w_1}  [Fig.~\ref{fig:spatial_w_2}]. From  Fig.~\ref{fig:spatial_w_1}, it is evident that the MF wavefunctions corresponding to the sector $w=1$ are localized on the side-coupled NM, with a small weight on KC. On the other hand, in the  sector $w=2$ [see Fig.~\ref{fig:spatial_w_2}], the MF wavefunctions have weights of similar magnitude on NM and KC.

\begin{figure*}[htb]
\includegraphics[width=5.3cm]{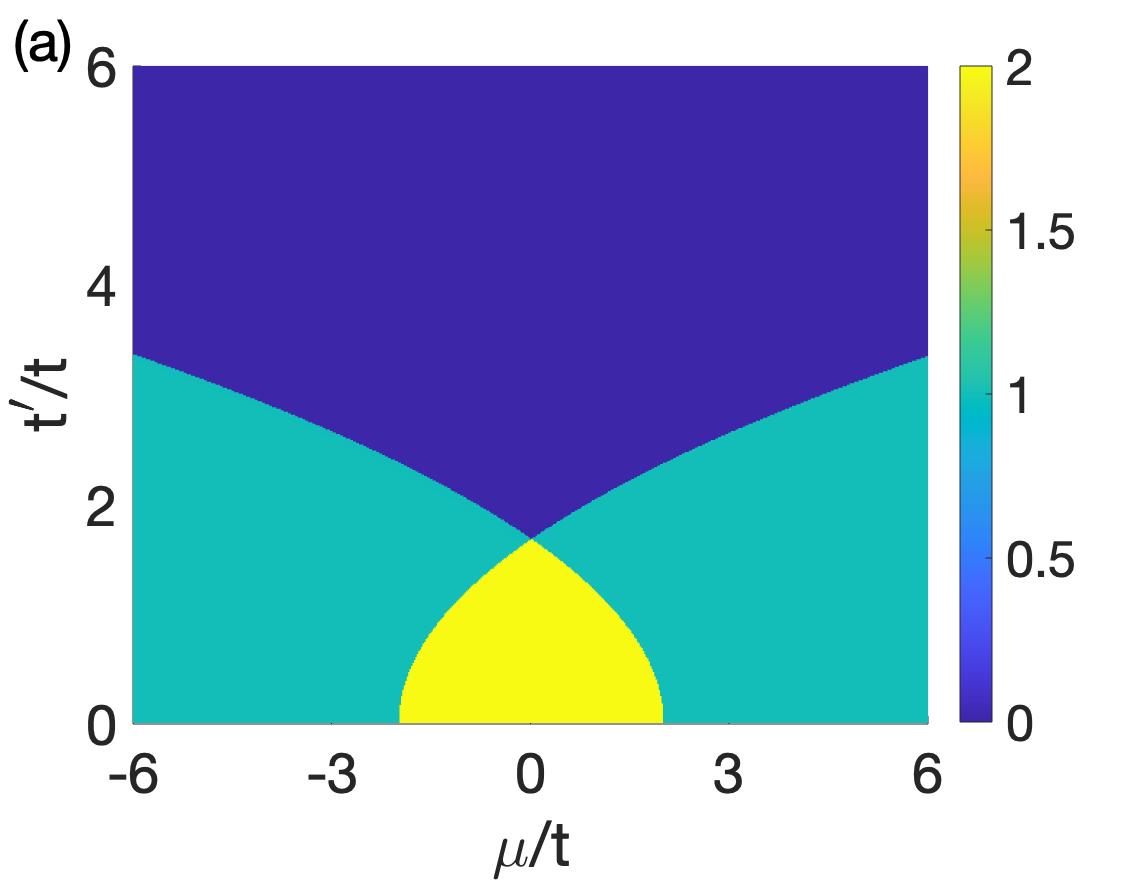}
\includegraphics[width=5.3cm]{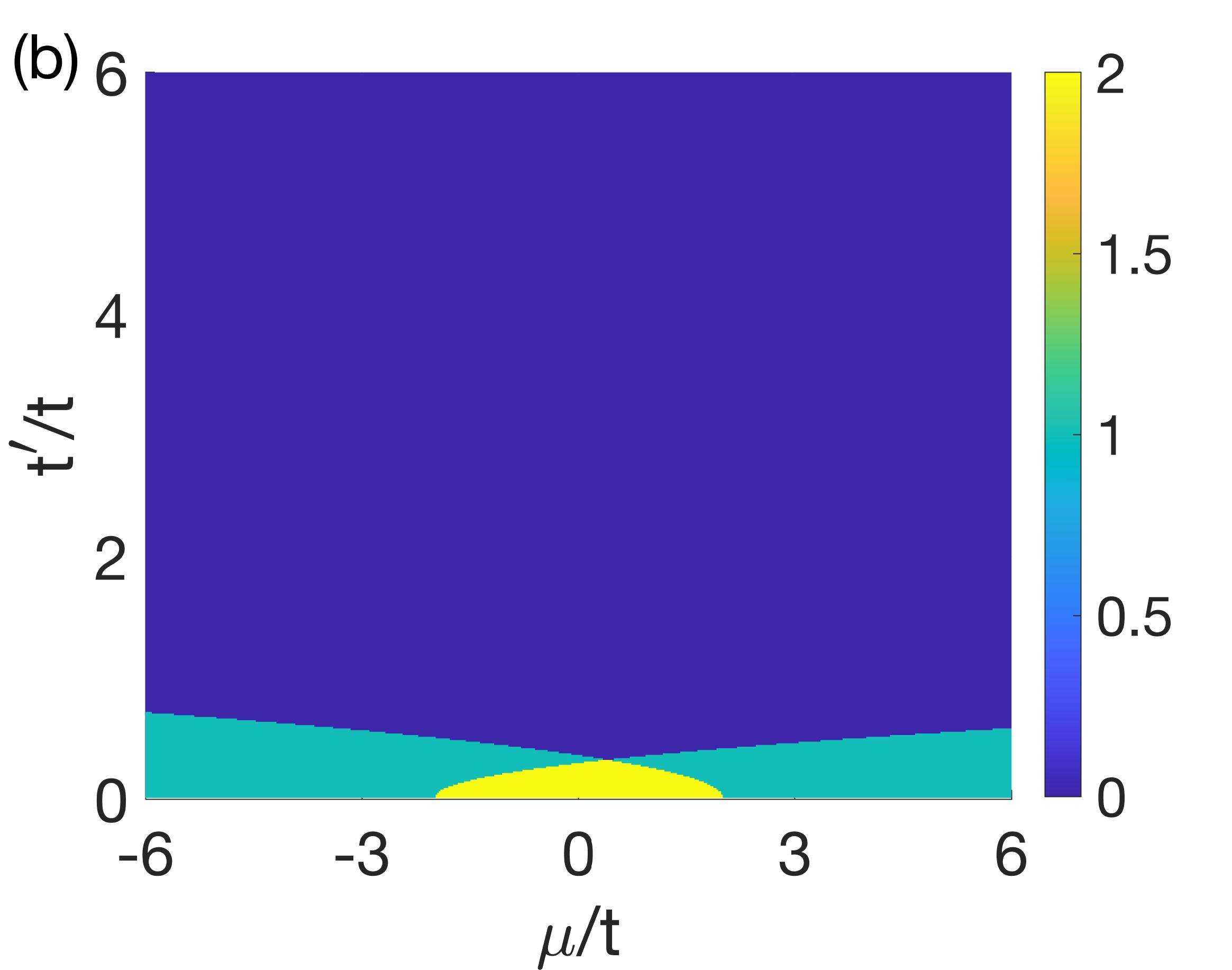}
\includegraphics[width=5.3cm]{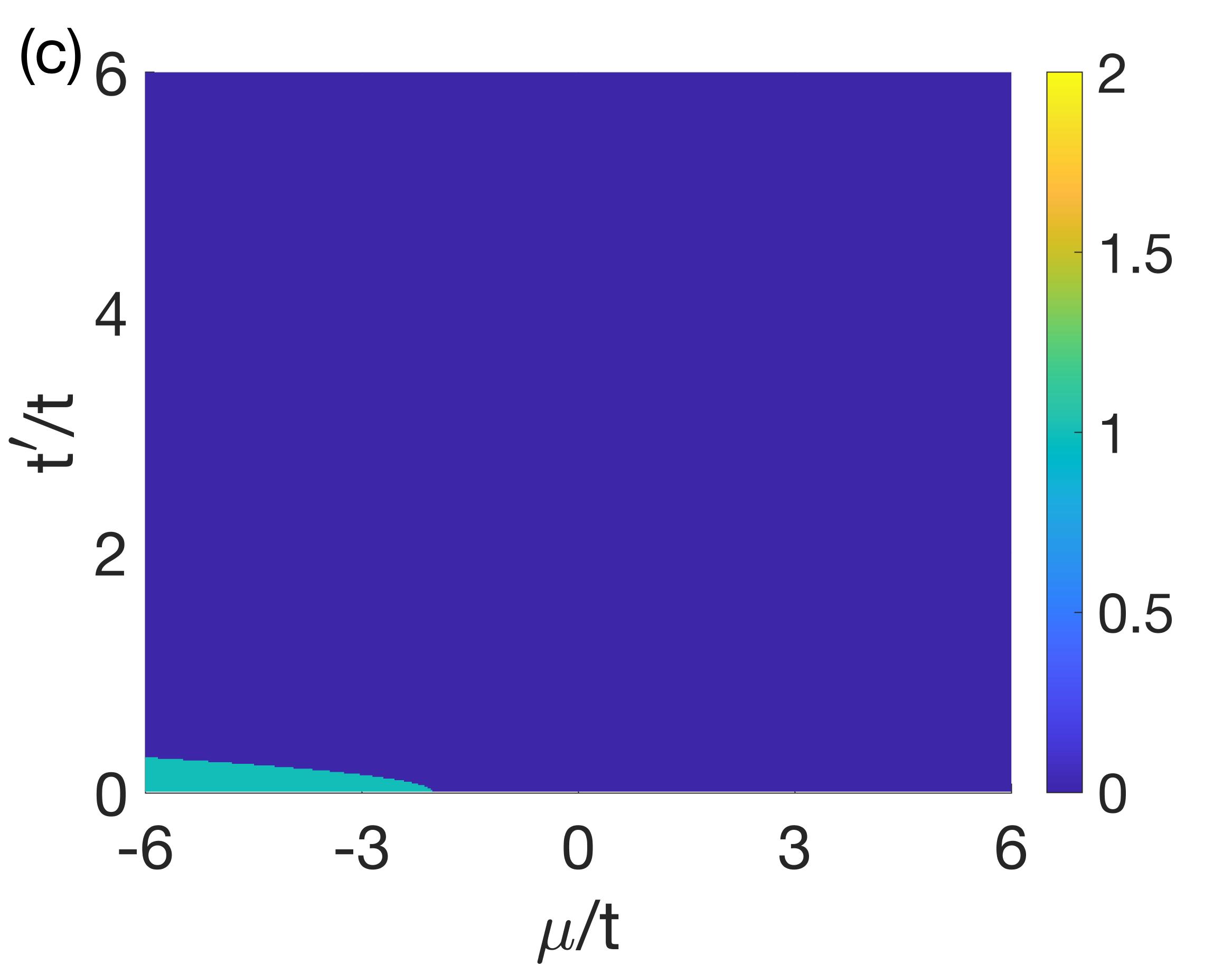}
\includegraphics[width=5.3cm]{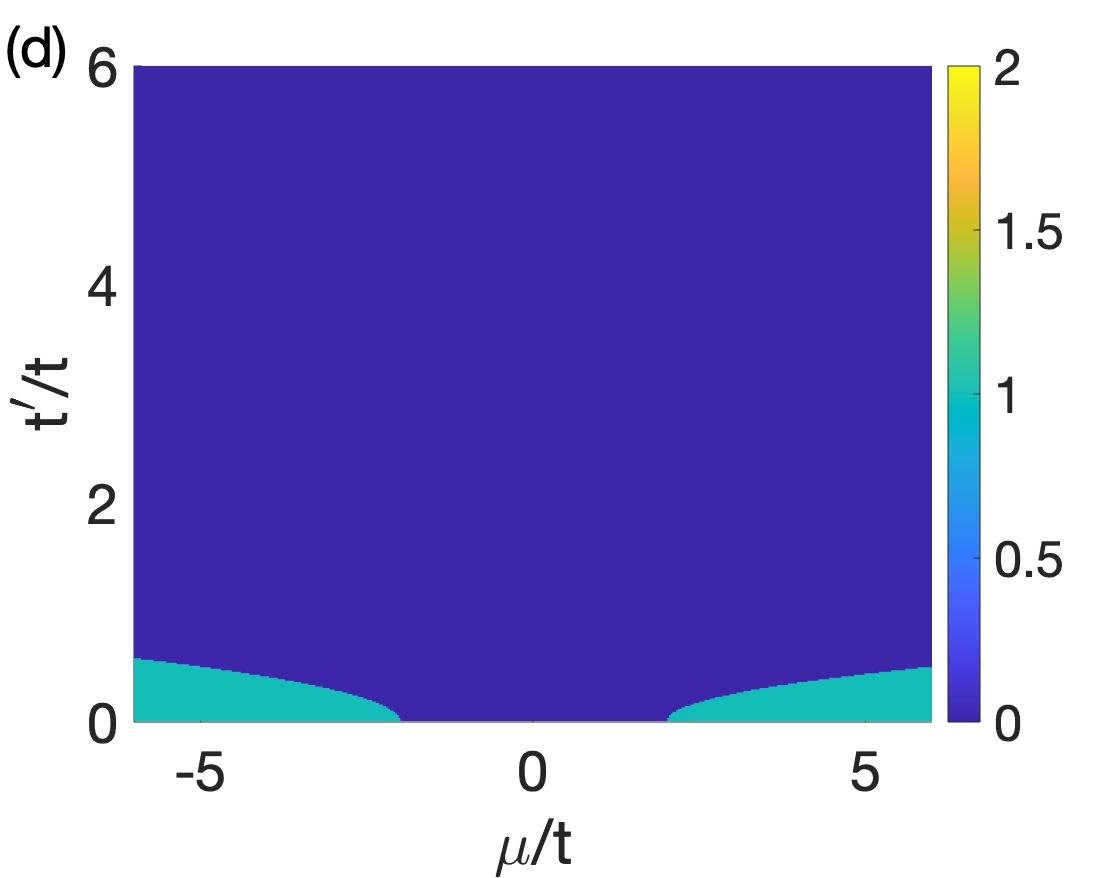}
\includegraphics[width=5.3cm]{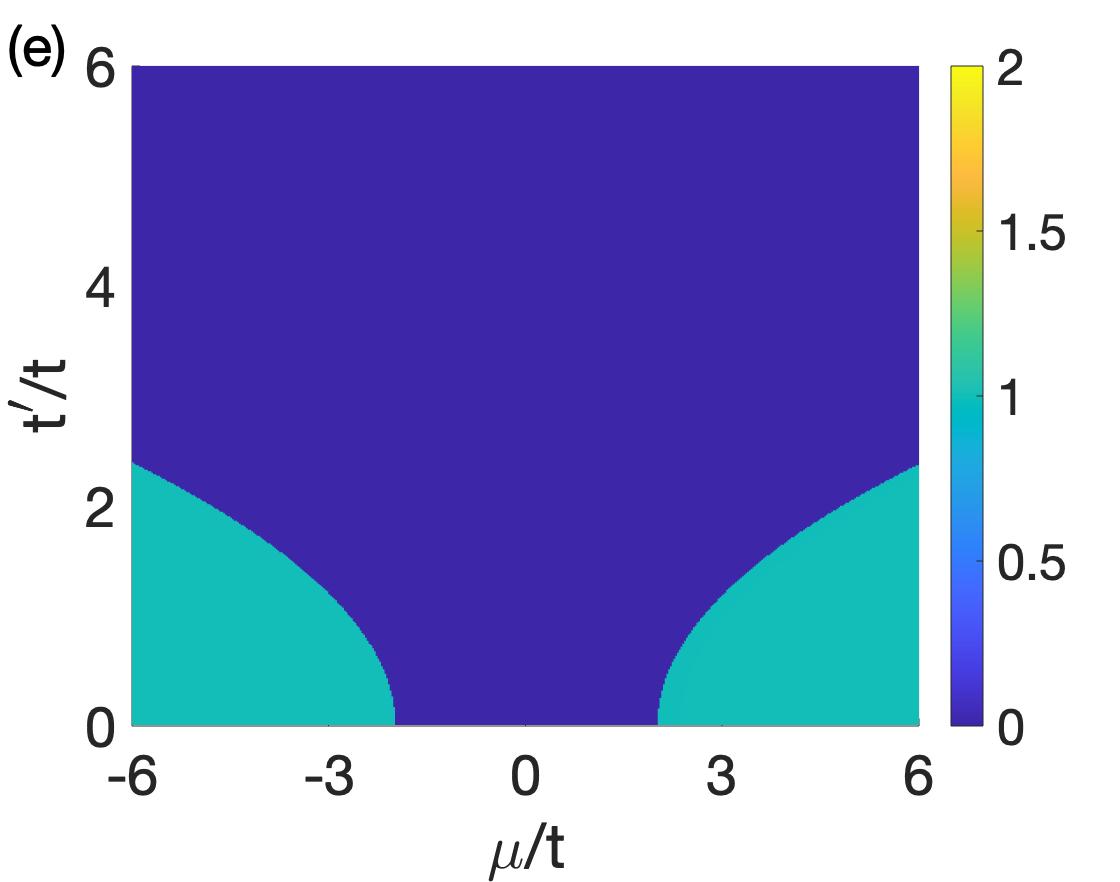}
\includegraphics[width=5.3cm]{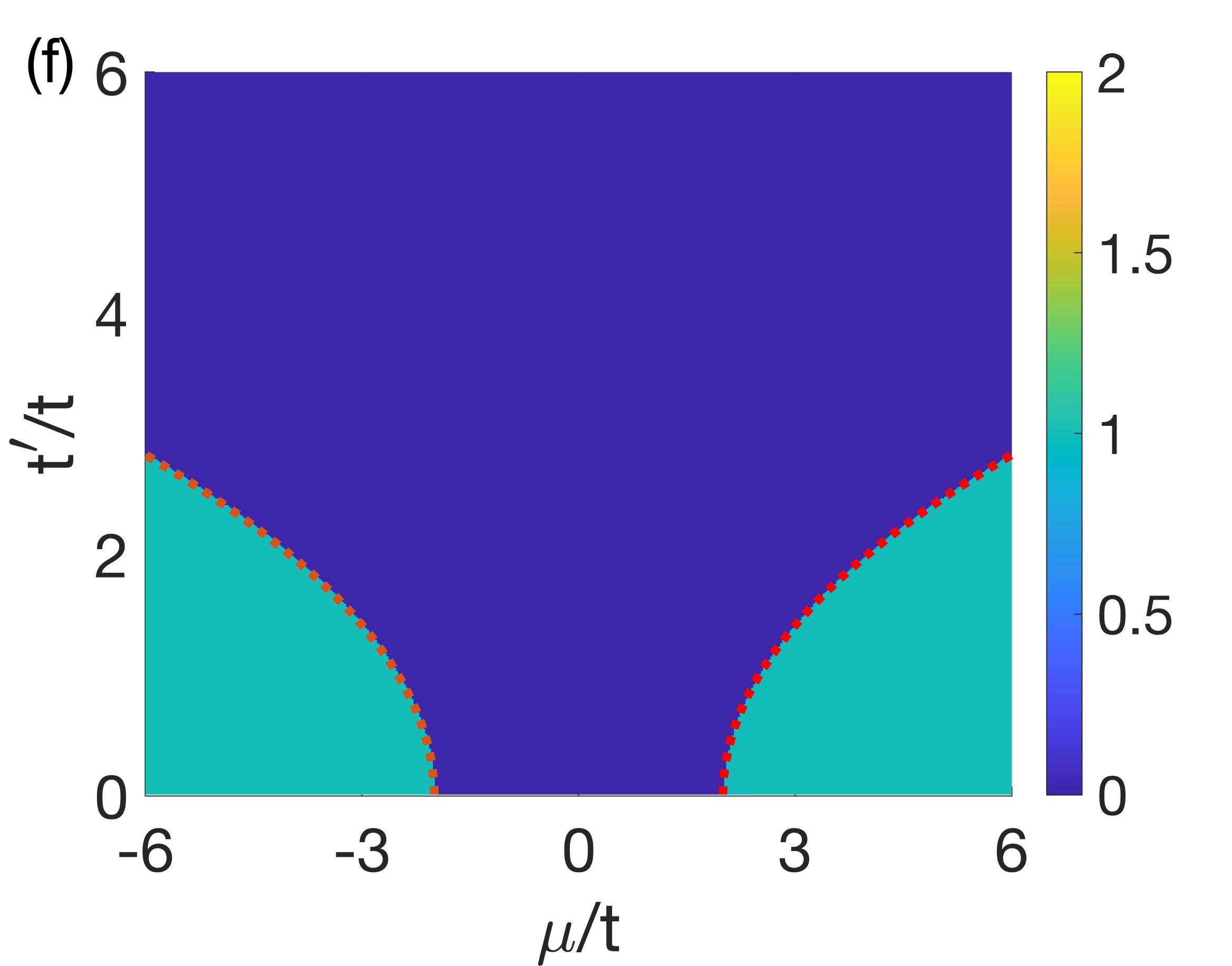}
\caption{Phase diagrams of the system for $\De=0.5t$, and different values of $\phi$. Winding number is plotted versus $t'/t$ and $\mu/t$.  (a) $\phi= 0.25\pi$, (b) $\phi=0.49\pi$, (c) $\phi=0.5\pi$, (d) $\phi=0.51\pi$, (e) $\phi=0.75\pi$, (f) $\phi=\pi$.  The data points with $t'=0$ are excluded. }\label{fig:winding-phi}
\end{figure*}

\subsection*{ Phase diagram for $\phi \neq 0$}
The procedure of calculating the winding number described in the previous section holds good even when $\phi\neq 0$. We calculate the winding number for different values of $\phi$ in Fig.~\ref{fig:winding-phi}. We find that the phase diagram changes as  $\phi$ is varied. As $\phi$ increases from $0$ to $\pi/2$, the region corresponding to $w=2$ in the phase diagram shrinks.  Interestingly, for the choice of $\phi\ge\pi/2$, there are only two phases: the one with single  pair of MFs and another with no MFs. We can understand the limiting case of $\phi=\pi$ by the following argument. The dispersion relations for the two electron bands of the Hamiltonian are $E=-\mu/2\pm\sqrt{t'^2+(\mu/2+2t\cos k)^2}$.  The superconductivity induces MFs if the electron bands cross $E=0$.  Of the two bands  here, only one band can cross $E=0$. It can be shown with some algebra that one band crosses $E=0$ when the parameters obey $|\mu|>2t+t'^2/2t$. The line separating the regions where this inequality is satisfied and not satisfied is plotted with red dotted line in Fig.~\ref{fig:winding-phi}(f) and the line coincides with the phase transition from $w=1$ to $w=0$.

\section{ Discussion and Conclusion}
The investigation into the behavior of MFs within the framework of the Kitaev model coupled to a one dimensional NM has yielded intriguing results. We began by analyzing the Hamiltonian governing the system, considering parameters such as hopping strength ($t$), chemical potential ($\mu$), pairing strength ($\Delta$), and coupling between the KC and NM ($t'$). Through numerical diagonalization, we explored the bulk spectrum and identified critical points where the gap in the energy spectrum closes. These critical points serve as indicators of phase transitions within the system. Further analysis revealed the presence of zero-energy states, which could potentially harbor MFs. By projecting the Hamiltonian onto different sectors defined by the electron and hole eigenstates, we calculated the winding number- a topological invariant- indicating the presence of MFs. The winding number analysis provided insights into the phase diagram of the system.  We find that the exact value of $\De$ does not change the phase diagrams as long as it is nonzero. Also, we have taken the chemical potential within the NM to be zero, thereby maintaining the NM in metallic phase. However, by hybridization with the KC, NM can be driven into band insulator phase when the pairing is turned off. An important message of this work is that even when KC and NM are in topologically trivial phase, coupling the two can result in a topological phase. 

Ref.~\cite{degottardi13prl,degottardi13} discuss the emergence of Majorna fermions in periodic potentials in a single Kitaev chain. They find that a finite pairing above a critical strength that depends on the amplitude of the periodic potential is required to get MFs in the system. The periodic potential on a NM drives it into insulating phase. So, it is interesting to see that MFs can be induced in such systems by a strong enough pairing strength. This is in contrast to our system where the coupled one dimensional system is driven into the topological phase hosting MFs only if the hybrid system is in metallic phase (in absence of pairing) and a small nonzero pairing strength on top of one chain is enough to get MFs. 

We extended our investigation to include the effect of magnetic flux ($\phi$) threading through the system. The phase diagram exhibited significant changes as $\phi$ varied, with the emergence of new phase transitions and alterations in the configuration of MFs. Notably, for $\phi\geq \pi/2$, the phase diagram simplified to two distinct phases-one with a single pair of MFs and another with none. Superconductivity repels magnetic fields. Manipulating MFs by using large magnetic fields has its own limitations, which can be overcome using small magnetic fluxes~\cite{lesser2021}. In this work, we find that enclosing small flux between NM and KC can help in manipulating the number of MFs. We can see from Fig.~\ref{fig:winding-phi} that small changes in magnetic flux alters the phase diagram significantly. This means, the setup studied in this work can have applications in sensing small magnetic fields. Since the magnetic flux is enclosed between NM and KC, the setup is different from multi-terminal Josephson junctions wherein the phase differences can be controlled by piercing a magnetic flux~\cite{heck2014}. Furthermore, having multiple pairs of MFs in parallel-connected chains can be advantageous for developing qubits, eliminating the need for braiding in gate operations~\cite{plugge2017}. Our setup, which hosts multiple phases and allows tuning of parameters to change the number of MF pairs, holds promise for applications in quantum computing.

\section*{ Acknowledgements }
The author thanks Diptiman Sen for useful comments on the work. 
The author thanks  Science and Engineering Research Board Core Research grant (CRG/2022/004311) for financial support. The author also thanks funding from University of Hyderabad Institute of Eminence PDF. 
\bibliography{ref_ksc}

\section*{ Appendix: Justification of the Winding Number Calculation Method }  A careful relook into the method used for calculating the winding number shows that when $\De$ is turned on, the two sectors $\si=\pm 1$ are not decoupled. This casts an element of doubt in the method followed. We justify the method used with the following sequence of arguments. Firstly, the matrix elements $\la\phi^e_{\si}|H_k|\phi^e_{\bar\si}\ra$ and $\la\phi^h_{\si}|H_k|\phi^h_{\bar\si}\ra$ (where $\bar\si=-\si$) are both zero. But, the matrix elements  $\la\phi^e_{\si}|H_k|\phi^h_{\bar\si}\ra$ and $\la\phi^h_{\si}|H_k|\phi^e_{\bar\si}\ra$ are non zero in general. This means, the coupling between the two sectors is such that electron states of one sector are coupled to hole states of the other sector. Also, at $k=0$, the pairing term in $H_k$ vanishes, making the two sectors decoupled. Secondly, for the formation of Majorana fermions, electron and hole excitations of the same sector need to form a linear superposition. Thirdly, we numerically  find that, the intra-sector pairing elements are much larger than the inter-sector pairing elements in regions of the phase diagram in Fig.~\ref{fig:winding}, where the winding number is $1$. For instance,  $\De_{inter}/\De_{intra}$  which quantifies the ratio of inter-sector pairing amplitude to the intra-sector pairing amplitude is found to be $0.1926$ for $t'=t$ and $\mu=5t$ (this ratio does not depend on the value of $\Delta$). Here,  $\De_{inter}$ is the maximum among all $k$ values in the range $(0,2\pi]$ of the larger value between $|\la\phi^e_{\si}|H_k|\phi^h_{\bar\si}\ra|$ and $|\la\phi^h_{\si}|H_k|\phi^e_{\bar\si}\ra|$, and $\De_{intra}$ is the maximum among all $k$ values in the range $(0,2\pi]$ of the larger between $|\la\phi^e_{\si}|H_k|\phi^h_{\si}\ra|$ and $|\la\phi^h_{\bar\si}|H_k|\phi^e_{\bar\si}\ra|$. The numerical finding that the ratio does not depend on the value of $\De$ implies that the two sectors can be treated separately even for large values of $\De$. Fourth, the gap closing curves in the spectrum shown in Fig.~\ref{fig:spectrum} do not change their position as $\De\neq 0$ is varied. Topological phase transitions are accompanied by gap closing. Since the gap closing lines do not change their position as $\De$ is varied, there is no topological phase transition upon tuning $\De\neq 0$.

\end{document}